\documentclass[12pt A4paper]{article}
\usepackage{latexsym}
\usepackage{graphicx}
\usepackage{amsmath}
\usepackage{amssymb}
\graphicspath{{./}{./figs/}}

\begin{document}
\title{Transient Resetting: A Novel Mechanism for Synchrony and Its Biological Examples \thanks{Citation: Li C, Chen L, Aihara K (2006) Transient resetting: A novel mechanism for synchrony and its biological examples. PLoS Comput Biol 2(8): e103. DOI: 10.1371/journal.
pcbi.0020103}}
\author{Chunguang Li$^{1,2,3}$, Luonan Chen$^4$, Kazuyuki Aihara$^{1,2}$}
%\email{aihara@sat.t.u-tokyo.ac.jp}
\date{\small $^1$ERATO Aihara Complexity Modelling Project, Room M204,
Komaba Open Laboratory, University of Tokyo,
4-6-1 Komaba, Meguro-ku, Tokyo 153-8505, Japan\\
$^2$Institute of Industrial Science, University of Tokyo, Tokyo 153-8505, Japan\\
$^3$ Centre for Nonlinear and Complex Systems, UESTC, Chengdu
610054, P. R. China\\
$^4$Department of Electrical Engineering and
Electronics, Osaka Sanyo University, Osaka, Japan} \maketitle

\begin{abstract}
The study of synchronization in biological systems is essential
for the understanding of the rhythmic phenomena of living
organisms at both molecular and cellular levels. In this paper, by
using simple dynamical systems theory, we present a novel
mechanism, named transient resetting, for the synchronization of
uncoupled biological oscillators with stimuli. This mechanism not
only can unify and extend many existing results on (deterministic
and stochastic) stimulus-induced synchrony, but also may actually
play an important role in biological rhythms. We argue that
transient resetting is a possible mechanism for the
synchronization in many biological organisms, which might also be
further used in medical therapy of rhythmic disorders. Examples on
the synchronization of neural and circadian oscillators are
presented to verify our hypothesis.
\end{abstract}

{\bf Synopsis:}

Synchronization of dynamical systems is a dynamical process
wherein two (or many) systems (either identical or nonidentical)
adjust a given property of their motions to a common behavior due
to coupling or forcing. Synchronization has attracted much
attentions of physicists, biologists, applied mathematicians, and
engineers for many years. In this paper, we present a very simple,
but generally applicable mechanism, named transient resetting, for
stimulus-induced synchronization of dynamic systems. The mechanism
is applicable not only to periodic oscillators but also to chaotic
ones, and not only to continuous time systems but also to discrete
time ones. Biological systems are dynamic, and the synchronization
in biological systems is essential, for example, for the
understanding of their rhythmic phenomena and information
processing. In this paper, we study several possible applications
in the biological context after presenting the novel mechanism. We
also show that transient resetting might also be used in medical
therapy of rhythmic disorders. Beneficial roles of noise in
biological systems have been extensively studied in recent years.
Our mechanism can also be seen as an explanation of the beneficial
roles of noise on the synchronization in biological systems,
though the stimulus is not necessarily required to be noisy in our
mechanism.

%{\bf Running Head:} Synchrony by Transient Resetting

\section*{Introduction}
Life is rhythmic. Winfree showed us a shocking discovery that a
stimulus of appropriate timing and duration can reset (stop) the
biological rhythm by driving the clock to a ``phase singularity",
at which all the phases of the cycle converge and the rhythm's
amplitude vanishes. He theoretically predicted this in the late
1960s, and then confirmed it experimentally for the circadian
rhythm of hatching in populations of fruitflies. Subsequent
studies have shown that mild perturbations can also quench other
kinds of biological oscillations, for example, the breathing
rhythms and neural pacemaker oscillations \cite{1, 2}. Such
findings may ultimately have medical relevance to disorders
involving the loss of a biological rhythm, such as sudden infant
death or certain types of cardiac arrhythmias \cite{3}. In
\cite{Tass}, Tass studied phase resetting by using methods and
strategies from synergetics. In \cite{4}, Leloup and Goldbeter
presented an explanation for this kind of long-term suppression of
circadian rhythms by the coexistence of a stable periodic
oscillation and a stable steady state in the bifurcation diagram.

On the other hand, synchronization is essential for biological
rhythms and information processing in biological organisms. So
far, many researchers have studied the synchronization in
biological systems experimentally, numerically and theoretically.
In this paper, by using simple dynamical systems theory, we show
that transient resetting, which concept will be clarified later,
can play a constructive role for biological synchrony. We argue
that transient resetting is a possible mechanism for synchrony
generally used in many biological organisms, which might also be
further used in medical therapy of rhythmic disorders. Winfree's
results showed the destructive aspect of the resetting, while we
show its constructive aspect in this paper. Although we
concentrate mainly on biological systems here, the novel mechanism
presented in this paper is applicable to general oscillators, so
in the following, we first present it as a general mechanism for
synchrony and then discuss its possible applications to biological
rhythms.

\section*{Results}
{\bf Basic mechanism}
\begin{figure}[htb]
\centering
\includegraphics[width=7cm]{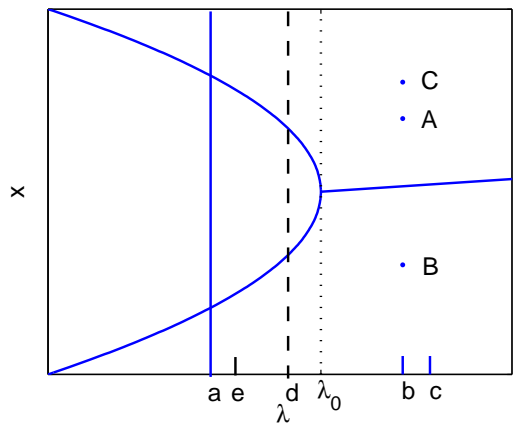}
\caption{A schematic bifurcation diagram with a normal
(supercritical) Hopf bifurcation for illustrating the transient
resetting mechanism.}
\end{figure}

Let's assume that an oscillator has a bifurcation diagram like
that shown in Fig. 1 with a normal (supercritical) Hopf
bifurcation \cite{5}, in which the bifurcation parameter $\lambda$
is the above mentioned ``stimulus'' and the curves with
$\lambda<\lambda_0$ show the maximum and minimum values of the
stable limit cycle. When the stimulus is a constant less than the
critical value (bifurcation point) $\lambda_0$, the system has a
stable rhythm (periodic oscillation); when the stimulus
$\lambda>\lambda_0$, the system has a stable equilibrium state,
which is shown by the curve with $\lambda>\lambda_0$. Undoubtedly,
many oscillators have this kind of bifurcation structure. Assume
that a population of (identical) oscillators operate in stable
periodic states with $\lambda =a$. To clarify the essential role
of transient resetting, we don't consider coupling among
oscillators in this paper. For oscillators without coupling, the
situation for studying the synchronization of two oscillators is
the same as that for a population of oscillators, so in the
following, we always consider two oscillators. Let's also assume
that there are some common fluctuations, e.g. periodic
fluctuations or random noise, that can perturb the parameter
$\lambda$ from $a$ to the right-hand side of $\lambda_0$, say to
$\lambda=b$ in Fig. 1, from time to time. When the duration for
the oscillators staying at the right-hand side of $\lambda_0$ is
long enough, the two oscillators will all converge to the steady
state, which is a rhythm-vanishing phenomenon. Next, let's examine
what will happen when the parameter $\lambda$ visit the value
$\lambda=b$ for a short duration. If the states of the two
oscillators are at points $A$ and $B$ respectively, the two
oscillators will have the tendency to converge to their common
steady state when the parameter $\lambda$ visits the value
$\lambda=b$, which means that the states of the two oscillators
will have the tendency to become closer. If the states of the two
oscillators are at the points $A$ and $C$ respectively, the two
oscillators will also have the tendency to converge to their
common steady state when the parameter $\lambda$ visits the value
$\lambda=b$. From conventional linear stability analysis, we know
that the velocity for converging to the steady state at point $C$
is higher than that at point $A$, so the two oscillators will also
have the tendency to become closer. Thus, by short-time visiting
the right-hand side of $\lambda_0$, the states of the two
oscillators always become closer, which is helpful for the genesis
of synchronization between the oscillators. Here the states of the
oscillators are not really reset to the steady state, but they
have the tendency (for a short time) to be reset to the steady
state, which is the reason we call it \emph{transient resetting}.

From the above analysis, it is easy to know that the bifurcation
is not necessarily required to be a supercritical Hopf
bifurcation. We only need that the oscillator operates, in most
times, in an oscillatory state, and can from time to time visit a
steady state in the parameter space (the solutions are all unique
in these two states). Even if there are other bifurcations between
these two states ($\lambda=a$ and $\lambda=b$), say a subcritical
Hopf bifurcation with coexistence of a stable limit cycle and a
stable equilibrium state between the dashed and the dotted
vertical lines in Fig. 1, the above argument can also hold.
Moreover, since oscillators are usually nonidentical in real
systems, there are some mismatches between the oscillators. If the
mismatches are not so large, however the oscillators can also be
synchronized (although not perfectly) by transient resetting. For
example, we assume that the mismatch can be reflected in the
parameter $\lambda$, and assume that the two oscillators operate
with parameters $\lambda=a$ and $\lambda=e$ respectively, and by a
common perturbation, the parameters of the two oscillators visit
$\lambda=b$ and $\lambda=c$ respectively in Fig. 1. If the
mismatch between the systems is not so large, the distance between
the steady states with $\lambda=b$ and $\lambda=c$ is most likely
small as well. The two oscillators have the tendency to be
contracted to the two steady states respectively, which mean that
roughly they are becoming closer since the two steady states are
close. In transient resetting, we don't care what the stimulus is.
It can be of any kinds, say periodic, random, impulsive, or even
chaotic stimuli, thus the transient resetting can unify many
existing results on stimulus-induced synchrony.

Next, we present several examples on biological rhythms to show
the effectiveness of the transient resetting and its biological
plausibility as a mechanism for biological synchrony.
\bigskip

{\bf Reliability of neural spike timing}

A remarkable reliability of spike timing of neocortical neurons
was experimentally observed in \cite{6}. In the experiments, rat
neocortical neurons are stimulated by input currents. When the
input is a constant current, a neuron generates different spike
trains in repeated experiments with the same input. It is evident
that the constant input when viewed as a bifurcation parameter has
moved the neuron dynamics from a steady state into a repetitive
spiking regime. It is shown that when a Gaussian white noise is
added to the constant current, the neuron generates almost the
same spike trains in repeated experiments. From the viewpoint of
synchronization, the repeated firing patterns imply that a common
synaptic current can induce almost complete synchronization in a
population of uncoupled identical neurons with different initial
conditions. This kind of synchronization may have great
significance in information transmission and processing in the
brain (see, e.g. \cite{7,8,9,10,11}).

We simulate the above mentioned behavior by using the well-known
Hodgkin-Huxley (HH) neuron model, which is described by the
following set of equations \cite{12}:
\begin{equation}
\begin{array}{rl}
C_m\dot{u}(t)=&G_{Na}m^3h(E_Na-u)+G_Kn^4(E_K-u)\\&+G_m(V_{\mbox{rest}}-u)+I_0+I(t),\\
\dot{m}(t)=&\alpha_m(u)(1-m)-\beta_m m,\\
\dot{h}(t)=&\alpha_h(u)(1-h)-\beta_h h,\\
\dot{n}(t)=&\alpha_n(u)(1-n)-\beta_n n,
\end{array}
\end{equation}
where $u(t)$ represents the membrane potential of the neural
oscillator, $m(t)$ and $h(t)$ the activation and the inactivation
of its sodium channel, $n(t)$ the activation of the potassium
channel, $I_0$ the constant input current, and $I(t)$ the
time-varying forcing. $\alpha_x$ and $\beta_x$ ($x=m,h,n$) are
rate functions that are given by the following equations:
\begin{equation*}
\begin{array}{c}
\alpha_m(u)=\frac{0.1(25-u)}{\mbox{exp}\left(\frac{25-u}{10}\right)-1},\,
\beta_m(u)=4 \mbox{exp}\left(-\frac{u}{18}\right),\\
\alpha_h(u)=0.07 \mbox{exp}\left(-\frac{u}{20}\right),\,
\beta_h(u)=\frac{1}{\mbox{exp}\left(\frac{30-u}{10}\right)+1},\\
\alpha_n(u)=\frac{0.01(10-u)}{\mbox{exp}\left(\frac{10-u}{10}\right)-1},\,
\beta_n(u)=0.125 \mbox{exp}\left(-\frac{u}{80}\right).
\end{array}
\end{equation*}
The parameters are set as the standard values \cite{12}, i.e.
$G_{Na}$=120 mS/cm$^2$, $E_{Na}$=115 mV, $G_K$=36 mS/cm$^2$,
$G_m$=0.3 mS/cm$^2$, $V_{\mbox{rest}}$=10.6 mV, and $C_m$=1
$\mu$F/cm$^2$. We always set the input $I_0+I(t)$ being zero for
$t<0$. When $I(t)=0$, if the value $I_0$ is larger than a critical
value $I_0^c\thickapprox 6$ $\mu$A/cm$^2$, the neuron has regular
spiking; otherwise, the neuron is in the steady resting state.
Thus, the bifurcation direction of HH neuron is opposite to that
shown in Fig. 1 and the type of Hopf bifurcation is subcritical
rather than supercritical. We fix the input constant $I_0=10$,
such that when $I(t)=0\, (t>0)$, this model exhibits a stable
periodic oscillation.

To simulate the experimental results in \cite{6}, we let
$I(t)=D\xi(t)$ in the HH model, where $D$ is a positive constant,
and $\xi(t)$ is a normal Gaussian white noise process with mean
zero and standard deviation one. We numerically calculate the
stochastic neuron model by using the Euler-Maruyama scheme
\cite{20} with time step $h=0.01$. When $D=2$, the two HH neural
oscillators can achieve complete synchronization as shown in Fig.
2 (a). One may ask whether it is really the ``noisy" property of
$I(t)$ that synchronizes the oscillators. We show here that this
behavior can be interpreted by the transient resetting mechanism.
In fact, when $I(t)$ is a random noise, it will make the total
input $I_0+I(t)$ visit the parameter region that is smaller than
the critical value $I_0^c\thickapprox 6$ from time to time, so
that the transient resetting mechanism can take effect. In the
experiments, the total input is indeed below the critical value in
some time durations, which implies that our argument is biological
plausible. The experimental (and simulation) results show that the
larger the noise intensity (to some extent), the higher the
reliability and precision of the spike timing, which can also be
interpreted by the transient resetting mechanism. In fact, when
the noise intensity is larger, the total input will spend more
time to be below $I_0^c$, such that the two trajectories will have
more opportunities to be contracted together. Due to the random
property of the input, when the input value is above $I_0^c$, the
effects of the noise for converging and diverging the two
trajectories are roughly balanced, thus, the longer the duration
of the input value being below $I_0^c$, the closer the two
trajectories and the faster the convergence should be. To further
verify the above argument that it is not the ``noisy" property of
the input that synchronizes the oscillators, we perform another
simulation, in which the fluctuation is a square wave with
amplitude of 4.5 and period of 20 ms, that is, the total input
switches between 10 (above $I_0^c$) and 5.5 (a little below
$I_0^c$) every 10 ms. In Fig. 2 (b), we show the simulation
result, in which the square wave is added at $t=80$. We see that
the two neural oscillators are synchronized rapidly, though the
input is a regular square wave.

\begin{figure}[htb]
\centering
\includegraphics[width=10cm]{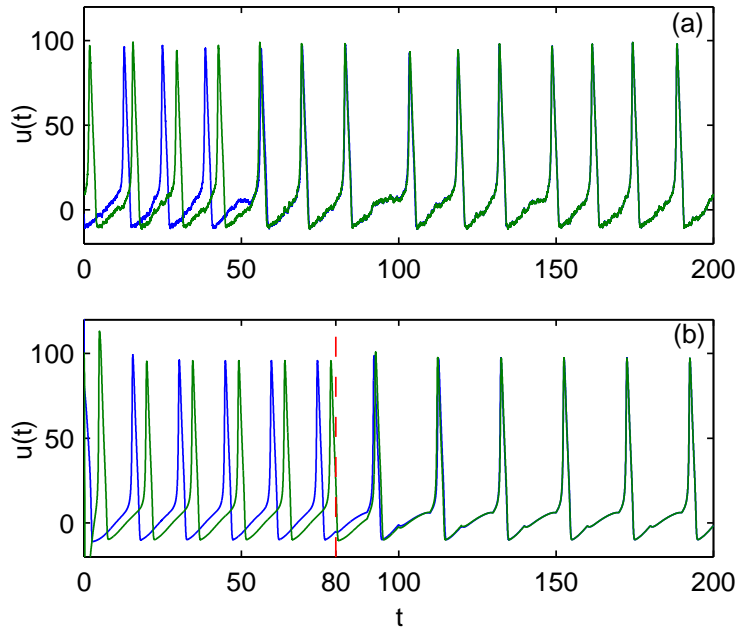}
\caption{Synchronization of two Hodgkin-Huxley neural oscillators:
(a) driven by common Gaussian white noise; and (b) driven by a
common square wave.}
\end{figure}

Next, we numerically study the relationship between the time rate
that the stimulus of the HH neuron model spends in the steady
state regime (defined as the {\bf duty}) and the time required to
achieve synchronization. For the convenience of measuring the
duty, we use square waves as the stimulus in the simulations. In
the following simulation, the period of the square wave is 20ms,
the time window when the stimulus is in the steady state parameter
regime is randomly chosen in each cycle, and all the other
parameters are the same as those in the above simulation. In Fig.
3, we plot the relationship between the duty and the time required
to achieve synchronization as the duty increases from 0.1 to 0.4
with step 0.05, in which the data are obtained by averaging the
results in 20 independent runs. Fig. 3 shows that the time
required to achieve synchronization decreases as the duty
increases. This quantitative result again confirms our above
argument and provides a numerical evidence that the proposed
mechanism can account for the observed synchrony.

\begin{figure}[htb]
\centering
\includegraphics[width=10cm]{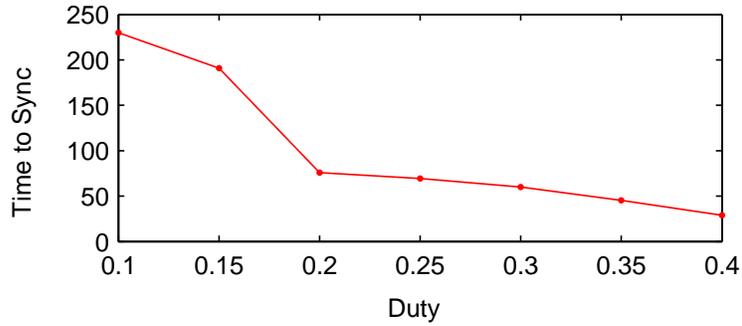}
\caption{The relationship between the duty and the time required
to achieve synchronization in the HH neuron model.}
\end{figure}

Biological oscillators are usually nonidentical in real systems,
and there are some mismatches between the oscillators. As we
mentioned in the Basic Mechanism section, if the mismatches are
not so large, the oscillators can also be synchronized (although
not perfectly) by the transient resetting. We next numerically
study the relationship between the parameter mismatch and the
synchronization error, that is, the robustness of the mechanism,
in the HH systems. In this simulation, we consider the mismatch on
$I_0$ in Eq. (1), namely in one of the two neural oscillators, the
constant input is $I_0-\Delta I$, and in the other one, it is
$I_0$. As we know, the input value $I$ has explicit effect on the
spiking frequency of the HH neuron model. We also use square wave
in this simulation, and the period and the duty are 10 and 0.3,
respectively. The synchronization error is defined as
$q=<|X_1(t)-X_2(t)|>$, where $X_1, X_2$ are the state vectors of
the two neural oscillators, that is
$X_i(t)=[u_i(t),m_i(t),h_i(t),n_i(t)]^T,i=1,2$, and $<\cdot>$ is
the average over time after discarding the initial phase of the
simulation for 100ms. In Fig. 4(a), we plot the relationship
between $\Delta I$ and $q$, in which each value of $q$ is also
obtained by averaging the results in 20 independent runs and the
error bars denote the standard deviations. Fig. 4(a) show that the
synchronization error $q$ increases with the increasing of $\Delta
I$, and when $\Delta I$ is not so large, the systems can also
achieve synchronization with a small synchronization error $q$. In
Fig. 4(b), we plot a typical simulation result with $\Delta
I=0.5$, and the $q$ value between 100ms and 200ms in this
simulation is $q=3.8667$. Fig. 4(b) demonstrates that the
transient resetting mechanism can indeed make the two nonidentical
neural oscillators synchrony, though the synchrony is not perfect.
When the mismatch becomes large, the systems may intermittently
lose synchrony, but shortly after that the systems can be
attracted back to synchrony again by the mechanism. When the
mismatch becomes much larger, the systems cannot maintain the
synchronous state, though the mechanism still have the tendency to
draw the systems together. In biological systems, the oscillators,
though not perfectly identical, can usually be similar and the
mismatch may not be so large. Thus the presented mechanism is
robust in such cases. Moreover, the synchronization in biological
systems may not be required to be perfect for emergence of
functions.

\begin{figure}[htb]
\centering
\includegraphics[width=10cm]{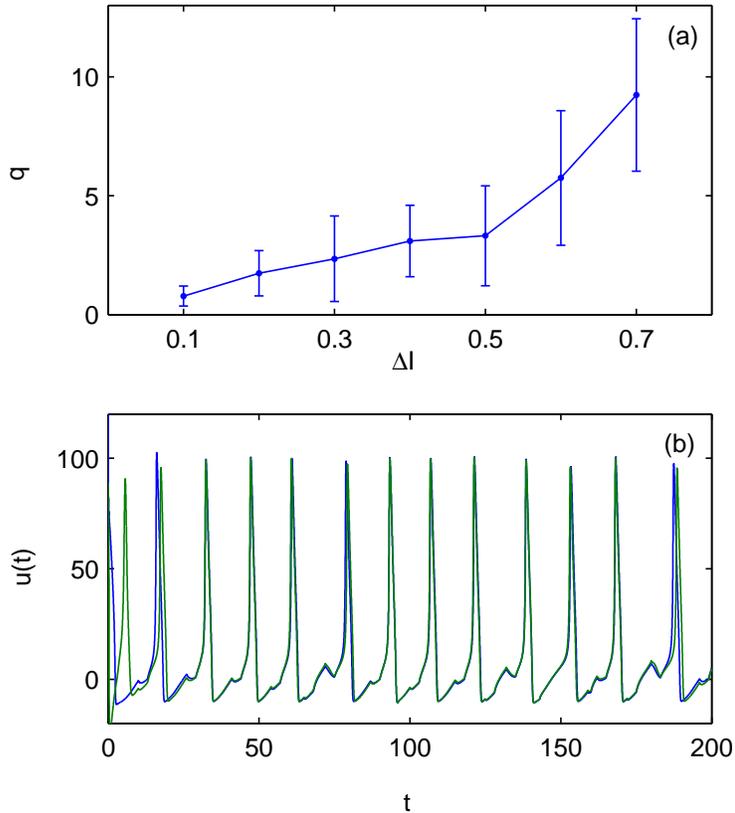}
\caption{Robustness of the mechanism with parameter mismatch on
$I_0$: (a) The relationship between the parameter mismatch $\Delta
I$ and the synchronization error $q$ in the HH neuron model. (b) A
typical numerical result of the systems with $\Delta I=0.5$.}
\end{figure}

It should be noted that when the mismatches exist in other
parameters, except the input, we can also obtain similar results.
For example, in another simulation, we consider a mismatch on the
parameter $G_m$, that is the parameter is $G_m+\Delta G_m$ in one
oscillator, and it is $G_m$ in the other oscillator. The
simulation method are the same as above. In Fig. 5(a), we show the
relationship between $\Delta G_m$ and $q$ as $\Delta G_m$
increases from 0.02 to 0.1 with step 0.02. In Fig. 5(b), we plot a
typical numerical result with $\Delta G_m=0.06$, and the $q$ value
between 100ms and 200ms in this simulation is $q=2.1492$. From
this figure, we can get the same conclusion as that in the above
example.
\begin{figure}[htb]
\centering
\includegraphics[width=10cm]{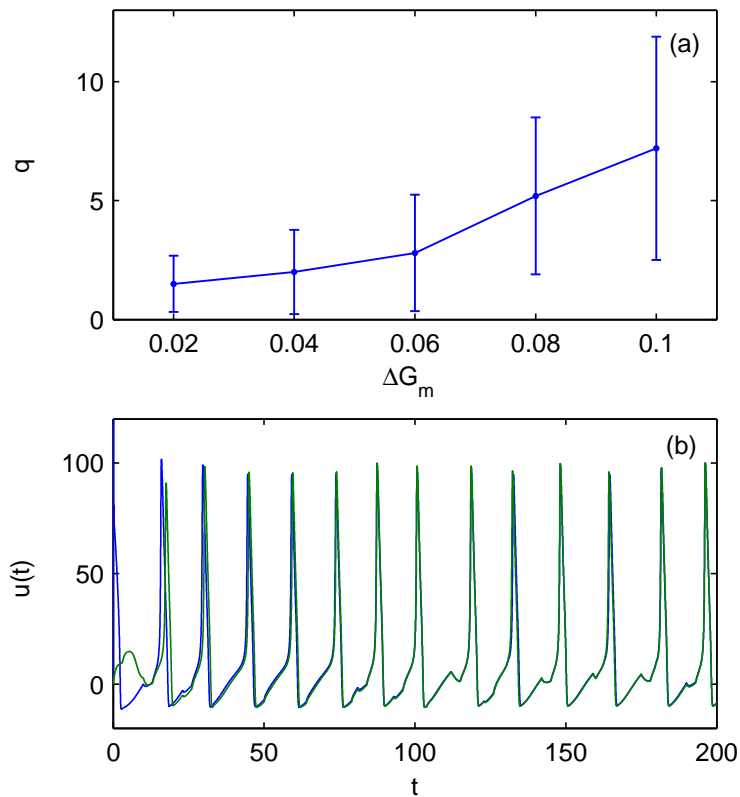}
\caption{Robustness of the mechanism with parameter mismatch on
$G_m$: (a) The relationship between the parameter mismatch $\Delta
G_m$ and the synchronization error $q$ in the HH neuron model. (b)
A typical numerical result of the systems with $\Delta G_m=0.06$.}
\end{figure}

It should also be noted that this transient resetting is effective
for both class I and class II neuron \cite{12}. Further, the Hopf
bifurcation of class II neurons can be either subcritical like the
HH model or supercritical associated with the canard phenomenon
\cite{13}.
\bigskip

{\bf Circadian oscillators}

In circadian systems, the light-dark cycle is the dominant
environmental synchronizer used to entrain the oscillators to the
geophysical 24-h day. In the following, we show that the
light-dark cycle as a synchronizer can also be interpreted by the
transient resetting mechanism. By our argument on transient
resetting, if the circadian clock has a bifurcation diagram
similar to Fig. 1 with a light-affected parameter as the
bifurcation parameter, and if the oscillators operate in the
oscillatory parameter region (say with $\lambda=a$ in Fig. 1) in
the dark duration, and in the steady state parameter region (say
with $\lambda=b$ in Fig. 1) in the light duration, the oscillators
may be automatically synchronized. For example, in the
Leloup-Goldbeter model of the {\it Drosophila} \cite{14}, the
light-dark cycle affects the degradation of the concentration of
TIM protein. In their parameter setting (which is biologically
plausible), under a continuous light condition, the model is in
the steady state parameter region. Then, according to the
transient resetting mechanism, the uncoupled oscillators can be
synchronized automatically, which is verified by our numerical
simulations (data not shown).

In theoretical studies of circadian clocks, the light-dark cycle
is usually represented by a square wave, but in fact, even with
continuous light, the light intensity includes fluctuations (light
noise). Next, we study the effect of light noise on the
synchronization of circadian oscillators. We consider the
Goldbeter circadian clock model of the {\it Drosophila} \cite{15}
as an example, which is described as follows:
\begin{equation}
\begin{array}{l}
\frac{dM}{dt}=v_s\frac{K_I^n}{K_I^n+P_N^n}-v_m\frac{M}{K_m+M},\\
\frac{dP_0}{dt}=k_sM-V_1\frac{P_0}{K_1+P_0}+V_2\frac{P_1}{K_2+P_1},\\
\frac{dP_1}{dt}=V_1\frac{P_0}{K_1+P_0}-V_2\frac{P_1}{K_2+P_1}-V_3\frac{P_1}{K_3+P_1}+V_4\frac{P_2}{K_4+P_2},\\
\frac{dP_2}{dt}=V_3\frac{P_1}{K_3+P_1}-V_4\frac{P_2}{K_4+P_2}-k_1P_2+k_2P_N-v_d\frac{P_2}{K_d+P_2},\\
\frac{dP_N}{dt}=k_1P_2-k_2P_N,
\end{array}
\end{equation}
where the parameter values are $v_s$=0.76 $\mu$Mh$^{-1}$,
$v_m$=0.75 $\mu$Mh$^{-1}$, $K_m=0.5$ $\mu$M, $k_s$=0.38 h$^{-1}$,
$v_d$=1 $\mu$Mh$^{-1}$, $k_1$=1.9h$^{-1}$, $k_2$=1.3h$^{-1}$,
$K_I=1$ $\mu$M, $K_d$=0.2 $\mu$M, $n$=4, $K_1=K_2=K_3=K_4=2$
$\mu$M, $V_1$=3.2 $\mu$Mh$^{-1}$, $V_2=1.58$ $\mu$Mh$^{-1}$,
$V_3$=5 $\mu$Mh$^{-1}$, and $V_4$=2.5 $\mu$Mh$^{-1}$ (see
\cite{15} for more details about this model). In this model, light
enhances the degradation of the PER protein by increasing the
value of $v_d$. With these biologically plausible parameter
values, the period of the circadian clock is about 24h. With the
increasing of $v_d$, there is a Hopf bifurcation similar to that
shown in Fig. 1, and the critical value of $v_d$ at the
bifurcation point is about 1.6. To clarify the effect of light
noise, we don't consider the light-dark cycle in this simulation,
that is $v_d=1$ is a constant (denoting the average light
intensity) if there is no light fluctuations. In our simulation,
we set $v_d=1+D\xi(t)$. When $D=0.15$, the simulation result is
shown in Fig. 6(a), which shows that the two circadian oscillators
achieve complete synchronization. This behavior can again be
interpreted by the transient resetting mechanism. The
interpretation is the same as that in the HH neural oscillator
case.

In the above examples, we show separately the effects of the
light-dark cycle and the light noise. The light-dark cycle itself
may not be strong enough to reach the steady state parameter
region in real biological circadian systems. For example, in Fig.
1, the light-dark cycle may make the bifurcation parameter switch
between $a$ and $d$ in the dark and the light durations,
respectively. In this case, a small light noise would drive the
parameter $\lambda$ to the right hand side of $\lambda_0$ from
time to time (in the light duration), which can be seen as the
combination or synergetic effect of the light-dark cycle and the
light noise. In other words, in biological circadian systems, it
is likely that the light-dark cycle and the light noise cooperate
to realize the transient resetting.
\begin{figure}[htb]
\centering
\includegraphics[width=10cm]{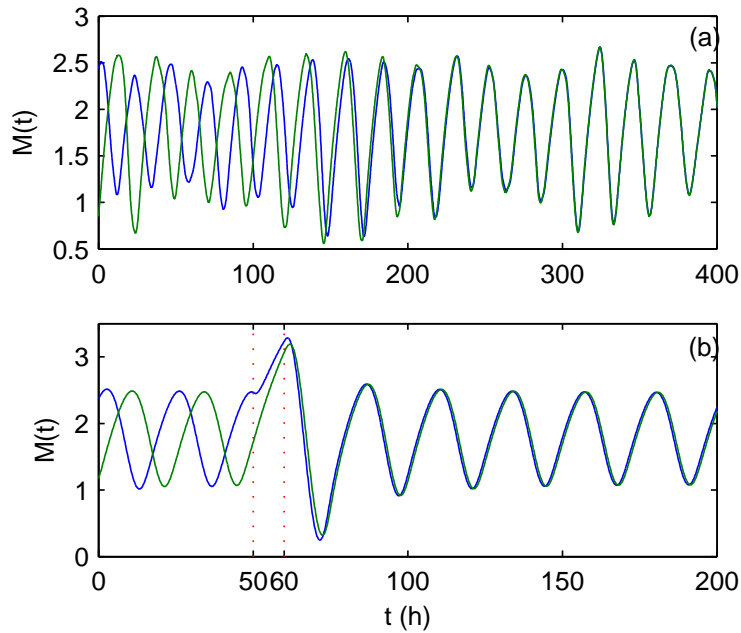}
\caption{Synchronization of two Goldbeter circadian oscillators:
(a) induced by common light noise; and (b) induced by a short
duration of bright light.}
\end{figure}

The time required to achieve synchrony and the robustness of the
mechanism in this system can also be studied similarly as in the
HH model, although we omit the detailed results here.
\bigskip

{\bf Therapy}

The transient resetting mechanism may have potential applications
in the therapy of various rhythmic disorders. Our analysis implies
that if we have some methods to control a biological rhythmic
system to make it visit its steady state parameter region
transiently, it may entrain the disordered rhythmic system to a
synchronous state. For example, we use a stimulation bright light
for 10 hrs with $v_d=1.7$ in $t\in [50, 60]$, and $v_d=1$ in other
time durations in the Goldbeter circadian oscillators. The
simulation result is shown in Fig. 6 (b), which shows that the two
oscillators are almost completely synchronized after the short
duration of the bright light stimulation. The exposure to bright
light also induces a several-hour delay shift of the circadian
oscillators, which is consistent with the experimental results
\cite{16}.

Except the above mentioned neural and circadian systems, the
mitotic control system \cite{17} may be another biological example
that uses transient resetting mechanism to achieve synchrony.
\bigskip

{\bf Chaotic neuron model}

In the above examples, we considered periodic oscillators, but the
transient resetting, as a general mechanism for synchrony, can
also be applied to chaotic systems. Clearly, the synchronization
of chaotic systems is more difficult, because uncoupled chaotic
oscillators, even with identical parameter values, will
exponentially diverge due to the high sensitivity to
perturbations. If the converging effect in steady state parameter
region is larger than the diverging effect in chaotic parameter
region, however the uncoupled chaotic systems may also be
synchronized by the mechanism. Here, we consider a simple
discrete-time chaotic neuron model described as follows \cite{21}:
\begin{equation}
x(t+1)=kx(t)-\alpha f(x(t))+a+I(t)
\end{equation}
where $f(x(t))=1/[1+\mbox{exp}(-x(t)/\epsilon)]$. This neuron
model is a model of the chaotic responses electrophysiologically
observed in squid giant axons \cite{22}. When decreasing $a$ from
a large to a small value, the system undergoes a period-doubling
road to chaos. We set the parameters $k=0.7,\alpha=1.05, a=0.93$
and $\epsilon=0.02$, such that when $I(t)=0$ the model exhibits
chaotic dynamics. When $I>0.11$, the neuron model is in a steady
state. In this example, we let $I(t)=0.15\xi(t)$ with $\xi(t)$ as
a normal Gaussian white noise process as defined before, such that
the neuron model can switch between chaotic states and steady
states. The numerical result in Fig. 7 indicates that the chaotic
neurons can indeed be synchronized by common noise, which can also
be interpreted by the transient resetting mechanism.
\begin{figure}[htb] \centering
\includegraphics[width=10cm]{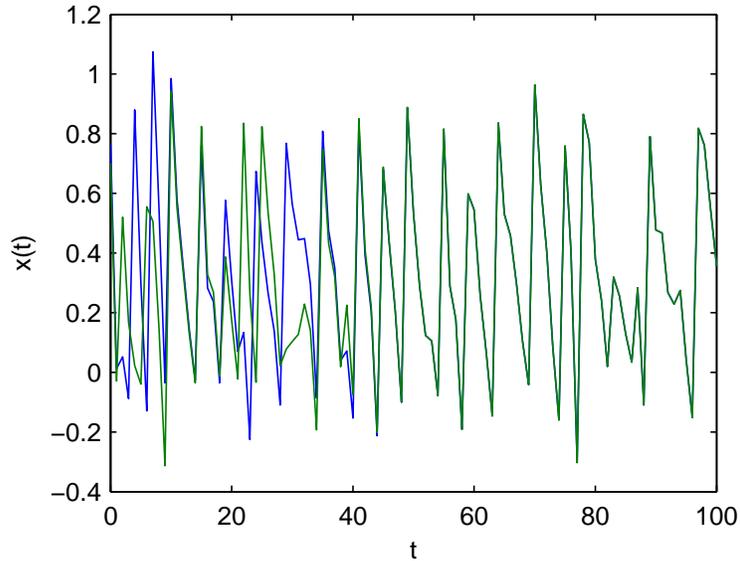}
\caption{Synchronization of two chaotic neurons induced by common
noise.}
\end{figure}

This example, though simple, shows that the transient resetting
mechanism is applicable not only to periodic oscillators but also
to chaotic systems, and not only to continuous time systems but
also to discrete time systems. We can also see that, as we
mentioned above, the bifurcation is not necessarily required to be
a Hopf bifurcation.
\section*{Discussion}
In literature, there exist some interesting results on
noise-induced synchronization of oscillators, which can be
classified into two classes: the noise depends or doesn't depend
on the states of the oscillators. In the case that that noise
depends on the states of the oscillators (see, e.g. \cite{18,
19}), the noise can, in fact, be seen as a kind of information
exchange or coupling with fluctuant coupling strengths, and it is
well known that coupling can induce synchronization, so it is not
surprising that the noise can induce synchronization. In the case
that noise doesn't depend on the states, many existing results can
be interpreted by the transient resetting proposed in this paper.
Some studies also theoretically explained the mechanisms of
periodic input forcing induced synchrony in literature. It should
be noted that the fluctuations in the present mechanism of
transient resetting are the parameters, not necessarily (but can
be) the inputs. Synchronization induced by the fluctuations of
some special parameters, for example time delay in delayed
systems, can also be interpreted by the mechanism. Thus the
transient resetting presented in this paper not only can unify and
extend many existing results on various fluctuation induced
synchrony, but also is very simple. It is reasonable to believe
that life systems, after long time of evolution, use as simple as
possible mechanisms to achieve complex functions.

It should also be noted, on the other hand, that although we have
shown in this paper the transient resetting as a possible general
mechanism for biological synchrony, we don't exclude other
possible mechanisms. Some systems driven by some {\it specific}
inputs, which don't satisfy the conditions shown in this paper,
can also be synchronized. It should not be surprising that in
biological systems, many mechanisms work together to jointly
guarantee the robustness and precision of synchrony.

In summary, in this paper, by simple dynamical systems theory, we
have presented a novel mechanism for synchrony based on the
transient resetting, and we have shown that it could be a possible
mechanism for biological synchrony, which can also potentially be
used for medical therapy. In contrast with Winfree's results, we
have shown the constructive aspect of (transient) resetting here.
In this paper, we are interested in the general qualitative
mechanism, so in the simulations, we didn't show many quantitative
details for each specific example. In Fig. 1, we showed a
1-parameter bifurcation diagram. In some systems, there might be
multiple parameters that affected by the fluctuations of stimuli.
In that case, we can use a similar multi-parameter bifurcation
diagram to understand the mechanism.

\section*{Materials and Methods}
To simulate the stocahstic differential equaiton
$\dot{x}(t)=f(x)+g(x)\xi(t)$, the Euler-Maruyama scheme is used in
this paper. In this scheme, the numerical trajectory is generated
by $x_{n+1}=x_n+hf(x_n)+\sqrt{h}g(x_n)\eta_n$, where $h$ is the
time step and $\eta_n$ is a discrete time Gaussian white noise
with $<\eta_n>=0$ and $<\eta_n\eta_m>=\delta_{nm}$. For more
details, see e.g. \cite{20}.

\section*{Acknowledgement}
The authors are grateful to the anonymous reviewers for their
valuable suggestions and comments, which have led to the
improvement of this paper. This work was partially supported by
Grant-in-Aid for Scientific Research on Priority Areas 17022012
from MEXT of Japan, the Fok Ying Tung Education Foudation under
Grant 101064, the National Natural Science Foundation of China
(NSFC) under Grant 60502009, and the Program for New Century
Excellent Talents in
University.\\
{\bf Conflicts of interest}. The authors have declared that no
conflicts of interest exist.\\
{\bf Author contributions}. CL conceived and designed the
numerical experiments, analyzed the data, contributed materials/
analysis tools. CL, LC and KA wrote the paper.

\end{document}